\documentclass[twocolumn,secnumarabic,amssymb, amsmath, nofootinbib,tightenlines,
nobibnotes, aps, prl,epsfig]{revtex4}
\usepackage{graphicx}
\usepackage{dcolumn}
\usepackage{bm}
\begin{document}
\preprint{APS/123-QED}
\title{Beyond DGLAP improved saturation model }

\author{G.R.Boroun}%
 \email{boroun@razi.ac.ir }
\affiliation{ Department of Physics, Razi University, Kermanshah
67149, Iran}
\date{\today}
\begin{abstract}
We present a modification of the DGLAP improved saturation model
with respect to the nonlinear correction (NLC). The GLR-MQ
improved saturation model is considered by employing the
parametrization of proton structure function due to the Laplace
transforms
 method, which preserves its
behavior success in the low and high $Q^{2}$ regions. We show that
the geometric scaling holds for the GLR-MQ improved model in a
wide kinematic region $rQ_{s}$. These results
 are comparable with other
 models in a wide kinematic region $rQ_{s}$. The behavior of the dipole cross sections,
 with respect to the GLR-MQ improved
saturation model, are comparable with the Color Glass Condensate
(CGC) model. The model describes the dipole cross sections in the
inclusive and diffractive processes. We also compare the nonlinear
corrections to the impact-parameter dependent saturation (IP-Sat)
model with the impact-parameter
dependent color glass condensate (b-CGC) dipole model. Finally, we consider the linear and nonlinear corrections to
 the IP Non-Sat model. These results provide a benchmark
for further investigation of QCD at small $x$ in future
experiments such as the Large Hadron Collider and Future Circular
Collider projects.\\

\end{abstract}
 \pacs{***}
\maketitle
\subsection{I. Introduction}

The color dipole picture (CDP) [1] has been introduced to study a
wide variety of small $x$ inclusive and diffractive processes at
HERA. The dipole approach, at small values of Bjorken $x$, gives a
clear interpretation of the high-energy interactions. This regime
of QCD is characterized by high gluon densities because the proton
structure is dominated by dense gluon systems [2-4] and predicts
that the small $x$ gluons in a hadron wavefunction should form a
Color Glass Condensate [5]. The gluon saturation effects are
observable at very small $x$ values and characterized by a hard
saturation momentum $Q_{s}(x)$. The saturation scale is a border
between dense and dilute gluonic systems as
\begin{eqnarray}
xg(x,Q_{s}^{2})\frac{\alpha_{s}(Q_{s}^{2})}{Q_{s}^{2}}{\simeq}{\pi}R^{2},
\end{eqnarray}
where $xg(x,Q^{2})$ is the gluon distribution function and
${\pi}R^{2}$ is the target area where $R$ is the correlation
radius between two interacting gluons. Indeed the parameter $R$
controls the strength of the nonlinearity. The saturation scale
rises with decreasing $x$ and at small enough $x$,
$Q_{s}{\gg}\Lambda_{\mathrm{QCD}}$ where $\Lambda_{\mathrm{QCD}}$
is the QCD cut-off parameter at each heavy quark mass threshold (i.e., $\Lambda^{n_{f}}_{\mathrm{QCD}}$).\\
Since nonlinear dynamics are known to become sizable only at
small-$x$, so the nonlinear contribution to the
Dokshitzer-Gribov-Lipatov-Altarelli-Parisi (DGLAP) evolution [6]
leads to an equation of the form
\begin{eqnarray}
\frac{\partial^{2}xg(x,Q^{2})}{\partial{\ln}(1/x)\partial{\ln}Q^{2}}=
\overline{\alpha}_{s}xg(x,Q^{2})-\frac{9}{16}\overline{\alpha}_{s}^{2}\pi^{2}
\frac{[xg(x,Q^{2})]^{2}}{{R}^{2}Q^{2}},
\end{eqnarray}
where $\overline{\alpha}_{s}{\equiv}{\alpha}_{s}C_{A}/\pi$ and the
value of $R$ is order of the proton radius
$(R\simeq5\hspace{0.1cm} \mathrm{GeV}^{-1})$, if the gluons are
distributed through the whole of proton, or much smaller
$(R\simeq2\hspace{0.1cm} \mathrm{GeV}^{-1})$ if gluons are
concentrated in hot spot within the proton. This was a vast
subject initiated by Gribov, Levin, Ryskin, Mueller and Qiu
(GLR-MQ) [7], as the second nonlinear term in (2) is responsible
for gluon recombination. This term arises from perturbative QCD
diagrams which couple four gluons to two gluons. So that two gluon
ladders recombine into a single gluon ladder. It leads to
saturation of the gluon density at low $Q^{2}$ with decreasing
$x$. A closer examination of the small $x$ scattering is
resummation powers of $\alpha_{s}\ln(1/x)$ where leads to the
$k_{T}$-factorization form [8]. In the $k_{T}$-factorization
approach the large logarithms $\ln(1/x)$ are relevant for the
unintegrated gluon density in a nonlinear equation.  Solution of
this equation develops a saturation scale where tame the gluon
density behavior at low values of $x$ and
this is an intrinsic characteristic of a dense gluon system [9].\\
The main goal of this paper is to consider the nonlinear
corrections to the DGLAP improved saturation model. In fact, the
DGLAP improved saturation model will be modified to the GLR-MQ
improved saturation model. This is based on the nonlinear
evolution of the gluon density at small values of $x$. These
results will be compared with the nonlinear saturation dynamics
which is explicitly incorporated into the CGC model. One of the
well-known impact-parameter dependent saturation models is the
IP-Sat model [10,11]. This is a simple dipole model that
incorporates the physics of saturation and all known properties of
the gluon saturation. In this case the saturation boundary is
approached via the DGLAP evolution, that is, by the eikonalization
of the gluon distribution, which effectively represents higher
twist contributions. The b-CGC and the IP-Sat models are easily
generalized from DIS off protons to DIS off nuclei [12].\\
This paper is organized as follows. In Sec. II, we introduce the
color dipole model for calculating the dipole cross sections in
the GBW, the DGLAP improved saturation, the b-CGC dipole , the
IP-Sat models and also the exclusive diffractive processes. In
Sec.III, we present the GLR-MQ improved saturation model to
consider the color dipole cross section at low values of $x$. Then
in Sec. IV, we present a detailed numerical analysis and our main
results. We summarize our main results in Sec.V.\\

\subsection{II. Dipole cross section}

Dipole representation provides a convenient description of DIS at
small $x$. There, the scattering between the virtual photon
$\gamma^{*}$ and the proton is seen as the
 color dipole where the transverse dipole size $r$ and the
 longitudinal momentum fraction $z$ with respect to the photon
 momentum are defined. The amplitude for the complete process is simply the production of
these subprocess amplitudes, as the DIS cross section is
factorized into a light-cone wave function and a dipole cross
section. Using the optical theorem, this leads to the following
expression for the $\gamma^{*}p$ cross-sections
\begin{eqnarray}
\sigma_{L,T}^{\gamma^{*}p}(x,Q^{2})=\int dz d^{2}\mathbf{r}
|\Psi_{L,T}(\mathbf{r},z,Q^{2})|^{2}\sigma_{\mathrm{dip}}({x},\mathbf{r}),
\end{eqnarray}
where subscripts $L$ and  $T$ referring to the transverse and
longitudinal polarization state of the exchanged boson. Here
$\Psi_{L,T}$ are the appropriate spin averaged light-cone wave
functions of the photon and $\sigma_{\mathrm{dip}}({x},r)$ is the
dipole cross-section which related to the imaginary part of the
$(q\overline{q})p$ forward scattering amplitude. The variable $z$,
with $0\leq z \leq 1 $, characterizes the distribution of the
momenta between quark and antiquark. The square of the photon wave
function describes the probability for the occurrence of a
$(q\overline{q})$ fluctuation of transverse size with respect to
the photon polarization [1, 2]. The dipole hadron cross section
$\sigma_{\mathrm{dip}}$ contains all information about the target
and the strong interaction physics with
\begin{eqnarray}
\sigma_{\mathrm{dip}}(x,r)=\int
d^{2}b\frac{d\sigma_{\mathrm{dip}}}{d^{2}b}
\end{eqnarray}
where $b$ is a particular impact parameter (IP) as
\begin{eqnarray}
\frac{d\sigma_{\mathrm{dip}}}{d^{2}b}=2(1-\mathrm{Re}~S(b)),
\end{eqnarray}
and $S(b)$ is the S-matrix element of the elastic scattering. The
cross section at a given impact parameter $b$ is proportional to
the dipole area, the strong coupling, the number of gluons in the
cloud and the shape function by the following form [10]
\begin{eqnarray}
\frac{d\sigma_{\mathrm{dip}}}{d^{2}b}=2\Big{[}1-
\exp\Big{(}-\frac{\pi^{2}r^{2}\alpha_{s}(\mu^{2})xg({x},\mu^{2})T(b)}{2N_{c}}\Big{)}
 \Big{]},
\end{eqnarray}
where the hard scale is  assumed to have the form
\begin{eqnarray}
\mu^{2}=C/r^{2}+\mu^{2}_{0},
\end{eqnarray}
and the parameters $C$ and $\mu^{2}_{0}$ are obtained from the fit
to the DIS data [1]. For multi Pomeron exchange, the eikonalised
dipole scattering amplitude of Eq.(6) can be expanded as
$$
N(x,r,b)=\sum_{n=1}^{\infty}\frac{(-1)^{n+1}}{n!}
\Big{[}\frac{\pi^{2}}{2N_{c}}r^{2}\alpha_{s}(\mu^{2})xg({x},\mu^{2})T(b)\Big{]}^{n}
,$$ where ${d\sigma_{\mathrm{dip}}}/{d^{2}b}=2N(x,r,b)$ and the
$n$-th term in the expansion corresponds to $n$-Pomeron exchange
[10]. Eq.(6) is known as the Glauber-Mueller dipole cross section
[13] and can  also be obtained within the McLerran-Venugopalan
model [14]. The exponential form  of the function $T(b)$ is
determined from the fit to the data as
\begin{eqnarray}
T(b)=\frac{1}{2{\pi}B_{G}}\exp(-b^{2}/2B_{G}),
\end{eqnarray}
where the parameter $B_{G}$ was found [10] to be
$4.25~\mathrm{GeV}^{-2}$.\\
In the original Golec-Biernat-W$\ddot{\mathrm{u}}$sthoff (GBW)
model [1], the dipole cross section was proposed to have the
eikonal-like form
\begin{eqnarray}
\sigma_{\mathrm{dip}}({x},r)=\sigma_{0}(1-e^{-r^{2}Q^{2}_{s}/4}),
\end{eqnarray}
where $Q_{s}({x})$ plays the role of the saturation momentum,
parametrized as $Q^{2}_{s}(x)=Q_{0}^{2}({x}/x_{0})^{-\lambda}$.
Parameters $Q_{0}$ and $x_{0}$ set dimension and absolute value of
the saturation scale and exponent $\lambda$ governs $x$ behavior
of $Q_{s}^{2}$. This model was updated in [2, 15] to improve the
large $Q^{2}$ description of the proton structure function by a
modification of the small $r$ behavior of the dipole cross section
to include the DGLAP evolved gluon distribution. Since the energy
dependence in large $Q^{2}$ region is mainly due to the behavior
of the dipole cross section at small dipole size $r$, therefore
authors in Refs.[2, 15] investigated the DGLAP evolution for small
dipoles. Bartels-Golec-Bienat-Kowalski (BGBK) improved the dipole
cross section by adding the collinear DGLAP effects. Indeed the
BGBK model is the implementation of QCD evolution in the dipole
cross section which depends on the gluon distribution. The
following modification of the DGLAP improved saturation model [1]
proposed for the dipole cross section as
\begin{eqnarray}
\sigma_{\mathrm{dip}}=\sigma_{0}\{1-\exp(-\frac{\pi^{2}r^{2}\alpha_{s}(\mu^{2})xg({x},\mu^{2})}{3\sigma_{0}})\}.
\end{eqnarray}
Indeed BGBK model is  successful in describing dipole cross
section at large values of $r$ as the two models (GBW and BGBK)
overlap in this region but they differ in the small $r$ region
where the running of the gluon distribution starts to play a
significant role. Indeed the DGLAP improved model of
$\sigma_{\mathrm{dip}}$ significantly improves agreement at large
values of $Q^{2}$ without affecting the physics of saturation
responsible for transition to small $Q^{2}$. As expected,
geometrical scaling is true for the DGLAP improved model curve for
the scaling variable $rQ_{s}{\geq}1$ and for the GBW model curve
for the whole region
[1].\\
The saturated version of the dipole model may in principle be
derived from the Color Glass Condensate effective theory for QCD
according to Eq.(6) where at small $r$ this expression (i.e.,
Eq.(6)) becomes
\begin{eqnarray}
\frac{d\sigma_{\mathrm{dip}}}{d^{2}b}=\frac{\pi^{2}r^{2}\alpha_{s}(\mu^{2})xg({x},\mu^{2})T(b)}{N_{c}}.
\end{eqnarray}
Eq.(6) is referred to as the IP-Sat model, while Eq. (11) is
referred to as the IP Non-Sat model. The BGBK  and CGC  models
considered only the dipole cross section integrated over the
impact parameter $b$ [16]. The BGBK model was modified to include
the impact parameter dependence as denoted by the IP-Sat model and
the CGC model was also modified to include the impact parameter
dependence as denoted by the b-CGC model. The dipole cross section
can be calculated in the CGC approach from the relation
\begin{eqnarray}
\sigma_{\mathrm{dip}}({x},r)=\sigma_{0}\mathcal{N}(x,r),
\end{eqnarray}
where $\sigma_{0}=2{\pi}R_{p}^{2}$ and
\begin{eqnarray}
\mathcal{N}(x,r)=\Big{\{}^{\mathcal{N}_{0}(\frac{rQ_{s}}{2})
^{2(\gamma_{s}+(1/k{\lambda}Y)\ln(2/rQ_{s}))}~:~rQ_{s}{\leq}2}
_{1-e^{-A\ln^{2}(BrQ_{s})}~~~~~~~~~~~~~~~~~:~rQ_{s}>2}
\end{eqnarray}
where $Y=\ln(1/x)$ and $k=\chi''(\gamma_{s})/\chi'(\gamma_{s})$
where $\chi$ is the LO BFKL [17] characteristic function. The
scattering amplitude $\mathcal{N}(x,r)$ can vary between zero and
one, where $\mathcal{N}=1$ is the unitarity limit. To introduce
the impact parameter dependence into the CGC model,  the b-CGC
model for the dipole cross section is defined by the following
form [16]
\begin{eqnarray}
\frac{d\sigma_{\mathrm{dip}}}{d^{2}b}&=&2\mathcal{N}(x,r,b)
\end{eqnarray}
where the impact parameter dependence of the saturation scale
$Q_{s}$ was introduced by
\begin{eqnarray}
Q_{s}{\equiv}Q_{s}(x,b)=(\frac{x_{0}}{x})^{\lambda/2}\Big{[}\exp(-\frac{b^{2}}
{2B_{CGC}})\Big{]}^{1/2\gamma_{s}},
\end{eqnarray}
where  the parameter $B_{CGC}$, instead of $\sigma_{0}$ in the CGC
dipole model, is a free parameter and is determined by other
reactions, namely the $t$ distribution of the exclusive
diffractive processes at HERA. The parameters were fixed by a
combination of theoretical
constraints [11] and a fit to DIS data.\\
Another one of the main advantages of dipole models is the
description of the diffractive process [2, 18]. The cross section
for the diffractive $q\overline{q}$ production reads [8]
\begin{eqnarray}
\frac{d\sigma_{L,T}^{D}}{dt}|_{t=0}=\int dz d^{2}\mathbf{r}
|\Psi_{L,T}(\mathbf{r},z)|^{2}\sigma^{2}_{\mathrm{dip}}({x},\mathbf{r}),
\end{eqnarray}
where $t=\Delta^{2}$, and $\Delta$ is the four-momentum
transferred into the diffractive system from the proton. In
Eq.(16), the generalised optical theorem is applied in the
framework of the dipole picture. At small values of the
diffractive mass $M^{2}\sim Q^{2}$ the elastic scattering of the
$q\overline{q}$ pair dominates, while at larger values of the mass
$M^{2}{\gg}Q^{2}$, the $q\overline{q}g$ contribution dominates
(due to gluon production in the final diffractive state). The
treatment of the $q\overline{q}g$ component goes beyond the
saturation model since this is not present in the inclusive
analysis [2, 18]. This component was computed in the two gluon
exchange approximation with an additional assumption of strong
ordering of transverse momenta of the $q\overline{q}$ pair and the
gluon. In the transverse coordinate representation, the
$q\overline{q}g$ system is treated as a color octet dipole
$8\overline{8}$ where the coupling of two $t$-channel gluons is
relative by a weight factor $C_{A}/C_{F}=2N_{C}^{2}/(N_{C}^{2}-1)$
with $C_{A}=N_{c}=3$ and
$C_{F}=\frac{N_{C}^{2}-1}{N_{C}}=\frac{4}{3}$ where $N_{C}$ is the
number of colors. Thus, the color dipole cross section for
exchange of a two gluon system for octet dipole reads [2,18]
\begin{eqnarray}
\sigma_{\mathrm{dip}}=\sigma_{0}\{1-\exp(-\frac{C_{A}}{C_{F}}\frac{\pi^{2}r^{2}\alpha_{s}(\mu^{2})xg({x},\mu^{2})}{3\sigma_{0}})\}.
\end{eqnarray}
In the next section, we consider the color dipole cross sections
due to the behavior of the linear and nonlinear gluon density and
compare with the other models. The linear gluon densities are
obtained with respect to the Laplace transform technique by
employing the parametrization of proton structure function, then
applied the GLR-MQ evolution equation  for the nonlinear gluon
densities. Some approximated analytical solutions in the color
dipole model have been reported in recent years [19, 20] with
considerable phenomenological success due to a parametrization of
the deep inelastic structure function for electromagnetic
scattering with protons.\\

\subsection{III. GLR-MQ improved saturation model}

 We will present an approach to the description of the color dipole cross section at small
 $x$, alternative to that based on the DGLAP improved saturation
 model. From a more theoretical viewpoint it is known that in
the low $x$, low $Q^{2}$ region  gluon recombination effects are
not negligible and reduce the growth of the gluon parton
distribution function. The GLR-MQ equation for the gluon density,
where an extra non-linear term, quadratic in the gluon density,
was added to the linear DGLAP evolution equation by the following
form
\begin{eqnarray}
\frac{{\partial}xg(x,\mu^{2})}{\partial{\ln}\mu^{2}}&=&
\frac{{\partial}xg(x,\mu^{2})}{\partial{\ln}\mu^{2}}|_{\mathrm{DGLAP}}\nonumber\\
&&-\frac{{81\alpha}_{s}^{2}}{16R^{2}\mu^{2}}
\int_{\chi}^{1}\frac{dy}{y}[yg(y,\mu^{2})]^{2},
\end{eqnarray}
where  $\chi=\frac{x}{x_{0}}$ and $x_{0}$ is the boundary
condition that the gluon distribution (i.e.,
$G(x,\mu^{2})=xg(x,\mu^{2})$) joints smoothly onto the linear
region. We note that at $x{\geq}x_{0}(=10^{-2})$ the non-linear
corrections are negligible.
 The nonlinear shadowing term, $\propto
-[g]^{2}$, arises from perturbative QCD diagrams. In this regime
the gluons in the proton form a dense system with mutual
interaction and recombination which also leads to the saturation
of the total cross section. Other early works on this topic can be
found in [21, 22].  In what follows, the hard scale is assumed to
have the form $\mu^{2}=C/r^{2}+\mu^{2}_{0} $ as for light quarks
the gluon distribution is evaluated at
$x=x_{BJ}=\mu^{2}/(\mu^{2}+W^{2})$ and for the charm quark the
gluon structure function is evaluated at
\begin{eqnarray}
x=(\mu^{2}+4m_{c}^{2})/(\mu^{2}+W^{2}),
\end{eqnarray}
  where $m_{c}$ is the
charm quark mass and $W$ refers to the photon-proton
center-of-mass energy. The non-linear equation (i.e., Eq.(18))
shows that the strong rise that is corresponding to the linear QCD
evolution equations at small-$x$ and $Q^{2}$ can be tamed by
screening effects. The first iteration of Eq.18 reads
\begin{eqnarray}
{d}G(x,\mu^{2})|_{\mathrm{NLC}}&=&
{d}G(x,\mu^{2})|_{\mathrm{DGLAP}}-\frac{{81\alpha}_{s}^{2}}{16R^{2}\mu^{2}}d{\ln}\mu^{2}\nonumber\\
&&{\times}\int_{\chi}^{1}\frac{dy}{y}[G(y,\mu^{2})]^{2},
\end{eqnarray}
where the nonlinear correction to the gluon distribution function
(i.e., $G^{\mathrm{NLC}}(x,\mu^{2})$ ) is obtained by the
following form
\begin{eqnarray}
G^{\mathrm{NLC}}(x,\mu^{2})&=&G^{\mathrm{NLC}}(x,\mu_{0}^{2})+[G(x,\mu^{2})-G(x,\mu_{0}^{2})]\\
&&-\int_{\mu_{0}^{2}}^{\mu^{2}}\frac{81}{16}\frac{\alpha_{s}^{2}(\mu^{2})}{{R}^{2}\mu^{2}}\int_{x}^{x_{0}}\frac{dz}{z}G^{2}(\frac{x}{z},\mu^{2})d{\ln}\mu^{2}.\nonumber
\end{eqnarray}
Here $G(x,\mu^{2})$ and $G(x,\mu_{0}^{2})$ are the linear gluon
distributions, and obtained from the parametrization $F_{2}$ using
the Laplace transform techniques [23, 24], at $\mu^{2}$ and
$\mu_{0}^{2}$ scales respectively. At the initial scale
$\mu_{0}^{2}$, the low $x$ behavior of the non-linear gluon
distribution is assumed to be [25]
\begin{eqnarray}
G^{\mathrm{NLC}}(x,\mu_{0}^{2})&=&G(x,\mu_{0}^{2})\{1+\frac{27\pi{\alpha_{s}(\mu_{0}^{2})}}{16{R}^{2}\mu_{0}^{2}}\theta(x_{0}-x)\nonumber\\
&&{\times}[G(x,\mu_{0}^{2})-G(x_{0},\mu_{0}^{2})] \}^{-1}.
\end{eqnarray}
Therefore the non-linear correction to the gluon distribution at
$\mu^{2}$ scale for $x<x_{0}$ reads
\begin{eqnarray}
G^{\mathrm{NLC}}(x,\mu^{2})&=&G(x,\mu^{2})+G(x,\mu_{0}^{2}){\Bigg{[}}\{1+\frac{27\pi{\alpha_{s}(\mu_{0}^{2})}}{16{R}^{2}\mu_{0}^{2}}\nonumber\\
&&{\times}[G(x,\mu_{0}^{2})-G(x_{0},\mu_{0}^{2})] \}^{-1}-1{\Bigg{]}}\\
&&-\int_{\mu_{0}^{2}}^{\mu^{2}}\frac{81}{16}\frac{\alpha_{s}^{2}(\mu^{2})}{{R}^{2}\mu^{2}}\int_{x}^{x_{0}}\frac{dz}{z}G^{2}(\frac{x}{z},\mu^{2})d{\ln}\mu^{2}.\nonumber
\end{eqnarray}
The gluon distribution due to the non-linear corrections can be
analytically solved at small $x$ with respect to the linear gluon
distribution behavior.\\
The linear gluon distributions (i.e., $G(x,\mu^{2})$ and
$G(x,\mu_{0}^{2})$) in Eq.(23) are defined with respect to the
most parametrization suggested in Refs. [23] and [24]. The authors
in Ref.[23] have an expression for the asymptotic part of $F_{2}$
(no-valence) as
\begin{eqnarray}
F_{2}{\propto}\ln^{2}(1/x)
\end{eqnarray}
for $x{\leq}0.09$. In Ref.[24], the authors obtained two quadratic
expressions  in $\ln^{2}(1/x)$ using second-order linear
differential equation as well as Laplace transforms for the
leading-order (LO) gluon distribution function, respectively. In
the first method in Ref.[24], the LO DGLAP equation for the
evolution of the proton structure function $F_{2}(x,Q^{2})$ is
rearranged into an inhomogeneous second-order differential
equation by the following form
\begin{eqnarray}
x^{2}\frac{\partial^{2}}{\partial{x^{2}}}G(x,Q^{2})-
2x\frac{\partial}{\partial{x}}G(x,Q^{2})+4G(x,Q^{2})=\nonumber\\
-\frac{4\pi}{\alpha_{s}}
\frac{9}{20}x^{4}\frac{\partial^{4}}{\partial{x^{3}}{\partial}{\ln}Q^{2}}\frac{F_{2}(x,Q^{2})}{x}
+\frac{12}{5}x\frac{\partial}{\partial{x}}F_{2}(x,Q^{2})\nonumber\\
-3x^{2}\frac{\partial^{2}}{\partial{x^{2}}}F_{2}(x,Q^{2})
-\frac{9}{5}x^{3}\frac{\partial^{3}}{\partial{x^{3}}}F_{2}(x,Q^{2})\nonumber\\
+\frac{12}{5}x^{4}\frac{\partial^{3}}{\partial{x^{3}}}
\int_{x}^{1}\frac{\partial}{\partial{x}}F_{2}(z,Q^{2}){\ln}\frac{z}{z-x}dz.
\end{eqnarray}
Eq.(25), with the new variable $\upsilon=\ln(1/x)$ becomes a
linear $2^{nd}$ order inhomogeneous equation, as
\begin{eqnarray}
{\bigg (}\frac{\partial^{2}}{\partial{\upsilon^{2}}}+
3\frac{\partial}{\partial{\upsilon}}+4 {\bigg
)}\widehat{G}(\upsilon,Q^{2})=
\mathcal{\widehat{G}}_{4}(\upsilon,Q^{2})
\end{eqnarray}
and the definition
$\widehat{G}(\upsilon,Q^{2})={G}(e^{-\upsilon},Q^{2})$. In
Ref.[24], the authors have found the parametrization of the gluon
distribution $\mathcal{\widehat{G}}_{4}(\upsilon,Q^{2})$
 which is calculated as a second degree polynomial in $\upsilon$ whose
coefficients are quadratic polynomials in $\ln(Q^{2})$ for
$x{\leq}0.09$ as
\begin{eqnarray}
\mathcal{\widehat{G}}_{4}(\upsilon,Q^{2})=\alpha(Q^{2})+\beta(Q^{2})\upsilon
+\gamma(Q^{2})\upsilon^{2},
\end{eqnarray}
Therefore
\begin{eqnarray}
\widehat{G}(\upsilon,Q^{2})=\frac{2}{\sqrt{7}}\int_{0}^{\upsilon}e^{-\frac{3}{2}(\upsilon-\upsilon')}
\sin(\frac{\sqrt{7}}{2}(\upsilon-\upsilon'))\mathcal{\widehat{G}}_{4}(\upsilon',Q^{2})d\upsilon',\nonumber
\end{eqnarray}
where the gluon distribution in $x$-space reads as a simple
quadratic polynomial in $\ln(1/x)$ with quadratic polynomial
coefficients in $\ln(Q^{2})$ by the following form
\begin{eqnarray}
G(x,Q^{2})=-0.459-0.143{\ln}Q^{2}-0.0155{\ln^{2}}Q^{2}+\nonumber\\
{\bigg [}0.231+0.00971{\ln}Q^{2}-0.0147{\ln^{2}}Q^{2}{\bigg
]}{\ln}(1/x)+\nonumber\\
{\bigg [}0.0836+0.06328{\ln}Q^{2}+0.0112{\ln^{2}}Q^{2}{\bigg
]}{\ln}^{2}(1/x).
\end{eqnarray}
In the second method, the authors [24] have suggested a new
parametrization based on Laplace transforms. The DGLAP evolution
is written as follows
\begin{eqnarray}
\int_{0}^{\upsilon}\widehat{G}(w,Q^{2})\widehat{h}(\upsilon-w)dw=\widehat{f}
(\upsilon,Q^{2}),
\end{eqnarray}
where $w=\ln(1/z)$ and
\begin{eqnarray}
\widehat{f}(\upsilon,Q^{2})=\frac{3}{4}\frac{4\pi}{\alpha_{s}}\mathcal{F}_{2}(e^{-\upsilon},Q^{2}).
\end{eqnarray}
The function $\widehat{h}(\upsilon)$ in Eq.(29) is
\begin{eqnarray}
\widehat{h}(\upsilon)=e^{-\upsilon}\widehat{{P}}_{gq}(\upsilon),
\end{eqnarray}
where $P_{gq}$ is the gluon-quark splitting function. The function
$\mathcal{F}_{2}(x,Q^{2})$ in Eq.(30) is sum of the proton
structure function ${F}_{2}$-dependent terms in the DGLAP
evolution equation by
\begin{eqnarray}
\mathcal{F}_{2}(x,Q^{2})=\frac{\partial{F_{2}(x,Q^{2})}}{\partial{\ln}Q^{2}}
-\frac{\alpha_{s}}{4\pi}{\Bigg
\{}\frac{16}{3}\int_{x}^{1}\frac{\partial{F_{2}(z,Q^{2})}}{\partial{z}}\nonumber\\
{\times}{\ln}\frac{z}{z-x}dz-\frac{4}{3}\int_{x}^{1}\frac{\partial{F_{2}(z,Q^{2})}}{\partial{z}}{\Bigg
( }\frac{x^{2}}{z^{2}}+\frac{2x}{z}{\Bigg )}dz{\Bigg \}}.
\end{eqnarray}
By making a Laplace transform in $\upsilon$, we can factor
Eq.(29), since the Laplace transform of a convolution is the
product of the Laplace transform of the factors, so that
\begin{eqnarray}
\mathcal{L}{\Bigg \{}
\int_{0}^{\upsilon}\widehat{G}(w,Q^{2})\widehat{h}(\upsilon-w)dw;s
{\Bigg \}}=\widehat{g}(s,Q^{2}){\times}\widehat{h}(s)
\end{eqnarray}
Solving Eq.(29) for g in $s$-space, we have
\begin{eqnarray}
\widehat{g}(s,Q^{2})=\widehat{h}^{-1}(s)\widehat{f}(s,Q^{2}).
\end{eqnarray}
Thus, inverting the Laplace transform of the factors, then the
gluon distribution is defined by
\begin{eqnarray}
\widehat{G}(\upsilon,Q^{2})=\mathcal{L}^{-1}\Big{[}\widehat{h}^{-1}(s)\widehat{f}(s,Q^{2});\upsilon
\Big{]}.
\end{eqnarray}
Therefore, the gluon distribution in $x$-space reads
\begin{eqnarray}
G(x,Q^{2})&=&\frac{9}{20}\frac{4\pi}{\alpha_{s}}\Big{\{}
3\mathcal{F}_{2}^{\gamma{p}}(x,Q^{2})-x\frac{\partial}{\partial{x}}\mathcal{F}_{2}^{\gamma{p}}(x,Q^{2})\nonumber\\
&&-\int_{x}^{1}\mathcal{F}_{2}^{\gamma{p}}(z,Q^{2})(\frac{x}{z})^{3/2}\Big{[}
\frac{6}{\sqrt{7}}\sin(\frac{\sqrt{7}}{2}\ln(\frac{z}{x}))\nonumber\\
&&+2\cos(\frac{\sqrt{7}}{2}\ln(\frac{z}{x}))\Big{]} \Big{\}},
\end{eqnarray}
for $0<x{\leq}0.06$. The standard representation for QCD coupling
in LO approximation is defined by
\begin{eqnarray}
\alpha_{s}^{\mathrm{LO}}(t)&=&\frac{4\pi}{\beta_{0}t},
\end{eqnarray}
where $\beta_{0}$ is the one loop correction to the QCD
$\beta$-function and $t=\ln\frac{Q^{2}}{\Lambda^{2}}$, $\Lambda$
is the QCD cut-off
parameter with $\alpha_{s}(M_{z}^{2})=0.118$.\\
The $\ln^{2}(1/x)$ behavior of the DIS proton structure function
(i.e., Eq.(24)) at small values of $x$ is compatible with
saturation of the Froissart bound at each value of $Q^{2}$. The
authors, in Ref.[23], have shown that this behavior may be the
signal for the saturation or gluon recombination processes  at
high parton densities. The gluon distribution in Eqs.(26-36),
according to the results in Ref.[24], is determined from the DGLAP
evolution equation for the proton structure function. Thus in
Eq.(20), the nonlinear corrections to the gluon behavior at low
$x$ and $Q^2$ values are considered, where it is compatible with
$\ln^{2}(1/x)$  behavior of parton densities at very small $x$ in
the
QCD evolution framework.\\
Now, we can estimate the non-linear corrections to the gluon
distribution (i.e., Eq.(23)) due to the linear gluon distributions
(i.e., Eq.(28) and (36)) for small $x$ and we will use the
non-linear corrections to the dipole cross sections, and in the
next section, the accuracy of the results will be discussed in comparison with the CGC model.\\


\subsection{IV. Results}

The linear and nonlinear methods are presented based on the
solutions of the DGLAP and  GLR-MQ evolution equations at the
leading-order accuracy in perturbative QCD, respectively. The
dipole cross-sections (Eqs.6, 10, 11 and 17) require the gluon
density $G(x,\mu^{2})$ for all scales $\mu^{2}$. These gluon
distributions [24] are obtained directly in terms of the
parameterization of the structure function $F_{2}(x,\mu^{2})$ and
its derivative. The resulting linear and nonlinear gluon
distribution functions for various dipole sizes for
$x{\leq}10^{-2}$ are shown in Fig. 1. The dipole size determines
the evolution scale $\mu^{2}$.
\begin{figure}[h]
\includegraphics[width=0.5\textwidth]{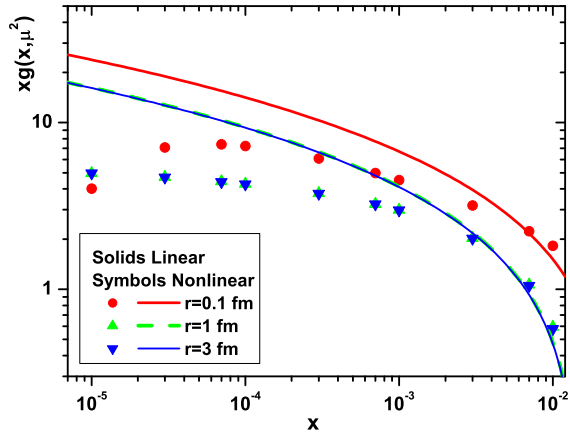}
\caption{The linear (Eqs.(28) and (36)) and nonlinear (Eq.(23))
gluon distribution functions for various dipole
sizes.}\label{Fig1}
\end{figure}
In this figure, we plot the $r$ dependence of the nonlinear
corrections to the gluon distribution for $R=2~\mathrm{GeV}^{-1}$.
Nonlinear corrections play an important role on gluon distribution
as $x$ and $\mu^{2}$ decrease. A depletion occurs at $x<10^{-2}$
where these results show that the nonlinear behavior of the gluon
distribution function is tamed. This taming behavior of nonlinear
gluon distribution function toward low $x$ at low $\mu^{2}$ values
becomes significant when considering the color dipole cross
section at the hot spot point (i.e., $R=2~\mathrm{GeV}^{-1}$). We
have calculated the linear and nonlinear corrections to the ratio
$\sigma_{\mathrm{dip}}/\sigma_{0}$ in a wide range of the dipole
size at the LO approximation. Results of calculations and
comparison with the GBW [1]  and CGC [5] models for $x=10^{-4}$
are presented in Fig.2. The linear corrections to the ratio of
color dipole cross sections at LO approximation are comparable
with the GBW model at low and high $r$ values. The nonlinear
corrections to the ratio of color dipole cross sections  are
comparable with the GBW model for $r{\lesssim}10^{-2}$ and $r
{\geq}10^{0}$ and also are comparable with the CGC model for
$10^{-2}{\lesssim}r{\leq} 10^{0}$. Indeed the nonlinear
corrections tame the behavior of the dipole cross section at
$r{\gtrsim}10^{-1}$. The effective parameters in the GBW model
have been extracted from a fit of the HERA data as,
$\lambda=0.288,~ x_{0}=3.04{\times}10^{-4},~ C=0.38$ and
$\mu_{0}^{2}=1.73$ [1]. Parameters of the CGC dipole model fixed
at the LO BFKL according to the original CGC fit [5] with respect
to  the values $\gamma_{s}=0.63$, $k=9.9$, $N_{0}=0.7$,
$\lambda=0.177$ and $x_{0}=2.70{\times}10^{-7}$ [16]. The dipole
cross sections are evaluated according to the four active flavors,
which take into account charm quark mass. The quark mass, in the
CGC model, was taken to be $1.4~\mathrm{GeV}$ although in our
calculations it is $1.29~\mathrm{GeV}$ [26].\\
\begin{figure}[h]
\includegraphics[width=0.55\textwidth]{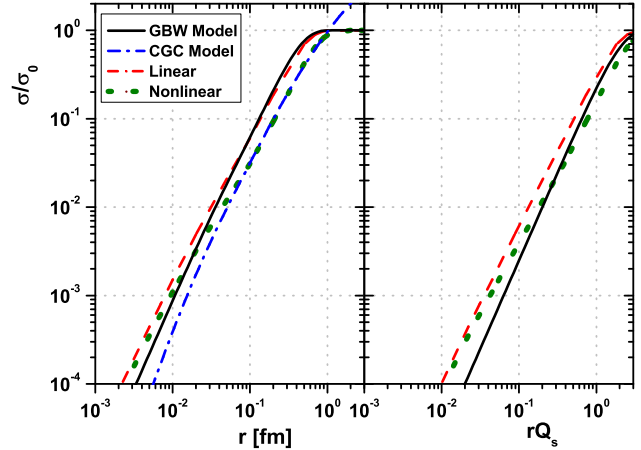}
\caption{Left:The extracted linear (dashed curve) and nonlinear
(dot curve) ratio $\sigma_{\mathrm{dip}}/\sigma_{0}$ (Eq.(10)) as
a function $r$ for $x=10^{-4}$ compared with the GBW model
(Eq.(9)) (solid curve) and CGC model (Eq.(13)) (dashed-dot curve,
CGC plotted due to Eq.(13) for $rQ_{s}{\leq}2$ and also the
parameters are defined from Ref.[16]). Right:The same as left as a
function $rQ_{s}$. }\label{Fig2}
\end{figure}
An important property of the saturation formalism is the geometric
scaling phenomenon, which means that the scattering amplitude and
corresponding cross sections can scale on the dimensionless scale
$rQ_{s}$. A particular interests present the linear and nonlinear
ratio $\sigma_{\mathrm{dip}}/\sigma_{0}$ defined by the scaling
variable $rQ_{s}$. In Fig.2 (right hand), we observe that the
nonlinear corrections to the  ratio
$\sigma_{\mathrm{dip}}(rQ_{s}(x))/\sigma_{0}$ merge into the GBW
curve for $rQ_{s}{{\gtrsim}}10^{-1}$. The results of the GLR-MQ
improved saturation model due to the parametrization of the proton
structure function have become a function of a single variable,
$rQ_{s}$, for almost all values of $r$ at LO approximation.\\
\begin{figure}[h]
\includegraphics[width=0.55\textwidth]{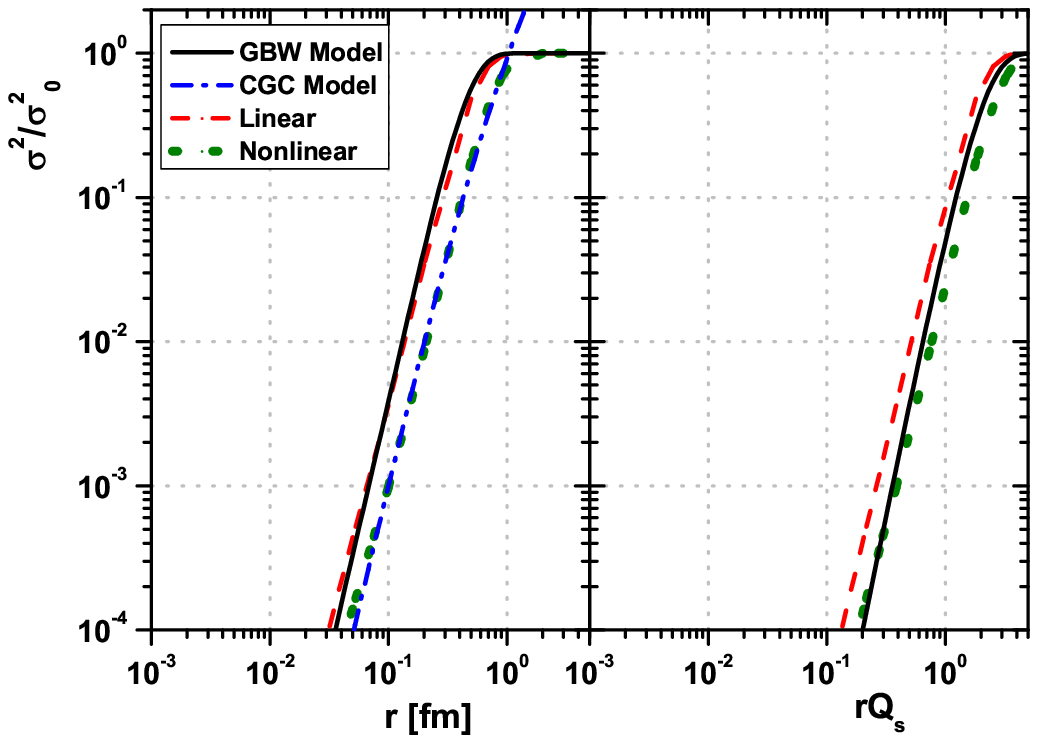}
\caption{The same as Fig.2 in the the simplest case of the
$q\overline{q}$ system for the ratio $\sigma
^{2}_{\mathrm{dip}}/\sigma^{2}_{0}$ (Eq.(16)). }\label{Fig3}
\end{figure}
The diffractive final state [2, 18] is built starting from a
$q\overline{q}$ pair in the color singlet state as the diffractive
$\gamma^{*}p{\rightarrow}q\overline{q}p'$ cross section is
proportional to  $\sigma^{2}(x,r)$ by Eq.(16). We have calculated
the ratio $\sigma ^{2}_{\mathrm{dip}}/\sigma^{2}_{0}$ for the
diffractive $q\overline{q}$ production into $r$ and $rQ_{s}$
respectively and compared the ratio with the GBW and CGC models
 in Fig.3. The linear corrections to the ratio $\sigma ^{2}_{\mathrm{dip}}/\sigma^{2}_{0}$
are comparable with the GBW model in a wide rang of $r$ although
the nonlinear corrections are comparable with the CGC model for
$r<1$ and with the GBW model for $r{\geq}1$. In Fig.3 (left hand),
the geometrical scaling of the nonlinear corrections to the ratio
$\sigma ^{2}_{\mathrm{dip}}/\sigma^{2}_{0}$ is visible in
comparison with the linear curve. The nonlinear curve merges into
one solid line in the right plot where the dipole cross section is
plotted as a function of the scaling variable $rQ_{s}$. This is a
reflection of geometric scaling in the nonlinear corrections in
comparison with the GBW model for the diffractive $q\overline{q}$
production.\\
\begin{figure}[h]
\includegraphics[width=0.55\textwidth]{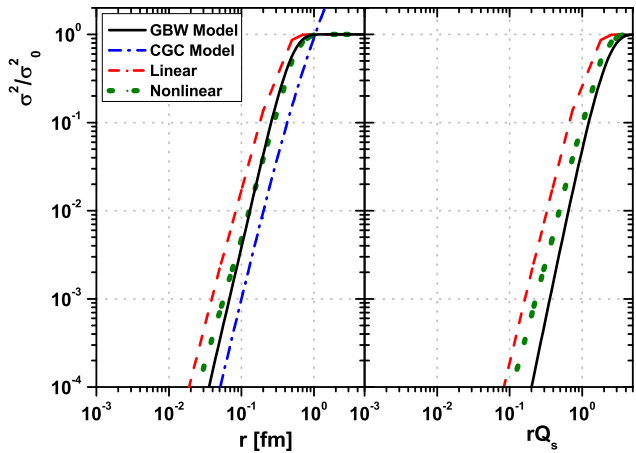}
\caption{The same as Fig.3 for the $q\overline{q}g$ component of
the diffractive system in the ratio $\sigma
^{2}_{\mathrm{dip}}/\sigma^{2}_{0}$ (Eq.(17)). }\label{Fig4}
\end{figure}
In addition to the contributions of the $q\overline{q}$ states, it
is important to include the contributions of the $q\overline{q}g$
final states of the diffractive processes in the nonlinear
corrections to the ratio $\sigma
^{2}_{\mathrm{dip}}/\sigma^{2}_{0}$. In Fig.4 the ratio of the
dipole cross sections are determined by the $q\overline{q}g$
component and are compared with the GBW and CGC models. The linear
and nonlinear ratio of the dipole cross sections, in comparison
with the results in Fig.3, are deviated from the GBW and CGC
models respectively. The reason for this deviation is because the
$q\overline{q}g$ component, interacting with the proton with the
same dipole cross section as the $q\overline{q}$ system, goes
beyond the saturation model [18]. Indeed, in Fig.4, the linear and
nonlinear cross sections are modified due to the weighted factor
$C_{A}/C_{F}$ although this component
is not present in the inclusive analysis.\\
\begin{figure}[h]
\includegraphics[width=0.55\textwidth]{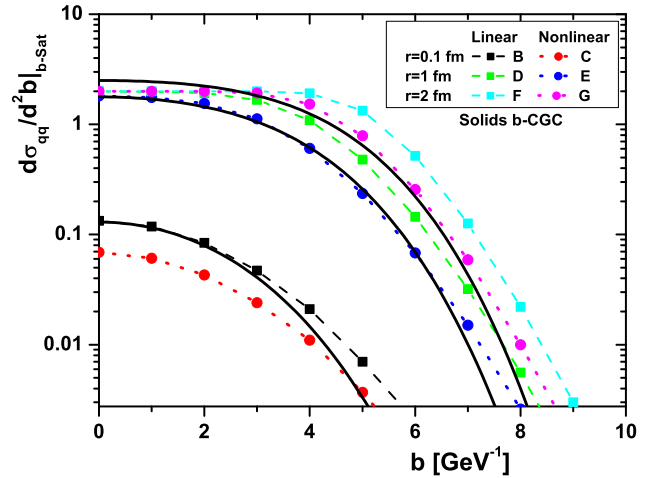}
\caption{The linear and nonlinear corrections to the impact
parameter dependent dipole cross section versus the impact
parameter $b$ (Eq.(6)) compared with the b-CGC model (Eq.(14)) for
the dipole sizes $r=0.1$, $1$ and $2~\mathrm{fm}$ at
$x=10^{-4}$.}\label{Fig5}
\end{figure}
Now we consider the nonlinear corrections to the the
$q\overline{q}$ differential cross section
$d\sigma_{\mathrm{dip}}/d^{2}b$. In Fig.5, the linear and
nonlinear corrections to the impact parameter dependent dipole
cross section due to the GLR-MQ equation are considered and
compared with the b-CGC model for $x=10^{-4}$. In this figure
(i.e., Fig.5) the linear corrections to the IP-Sat (i.e., b-Sat)
model are comparable with the b-CGC model in a wide range of the
impact parameter $b$ for $r<1~\mathrm{fm}$ and the nonlinear
corrections to the IP-Sat model are comparable with the b-CGC
model in a wide range of the impact parameter $b$ for
$r{\geq}1~\mathrm{fm}$. The optimum values for the b-CGC model
parameters are the following [16]: $\gamma_{s}=0.46$,
$B_{CGC}=7.5~\mathrm{GeV}^{-1}$, $N_{0}=0.558$,
$x_{0}=1.84{\times}10^{-6}$ and $\lambda=0.119$. In Fig.5 we
observe that the linear and nonlinear behavior of
$d\sigma_{\mathrm{dip}}/d^{2}b$ grows rapidly with $r$ for small
values of $b$, until those reach the saturation plateau,
$d\sigma_{\mathrm{dip}}/d^{2}b=2$, which illustrates saturation in
the Glauber- Mueller approach. Indeed, the GLR-MQ improved
saturation model illustrates unitarity with an increase of $r$ as
$b$ decreases.\\
\begin{figure}[h]
\includegraphics[width=0.55\textwidth]{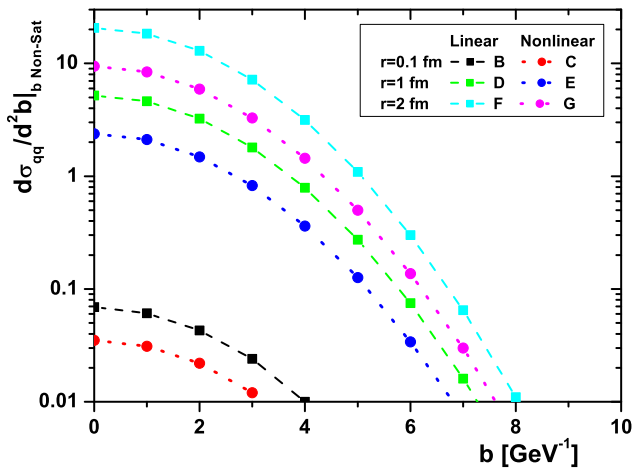}
\caption{The linear and nonlinear corrections to the IP Non-Sat
(Eq.(11)) versus the impact parameter $b$ for the dipole sizes
$r=0.1$, $1$ and $2~\mathrm{fm}$ at $x=10^{-4}$.}\label{Fig6}
\end{figure}
At small $r$, the IP-Sat model (Eq.6) becomes the IP Non-Sat model
(Eq.11) where the interaction between the dipole and the hadron is
described by the exchange of one gluon. The linear and nonlinear
behavior of $d\sigma_{\mathrm{dip}}/d^{2}b$ in the IP Non-sat
model are considered in Fig.6. In this model, the behavior of the
$d\sigma_{\mathrm{dip}}/d^{2}b$ is directly dependent on the gluon
distribution function. Saturation effects are not visible in this
model as $b$ decreases. However this behavior tamed due to the
nonlinear corrections to the gluon density. For small dipole sizes
the distributions are almost similar but they differ significantly
as $r$ becomes large.\\
\begin{figure}
\includegraphics[width=0.55\textwidth]{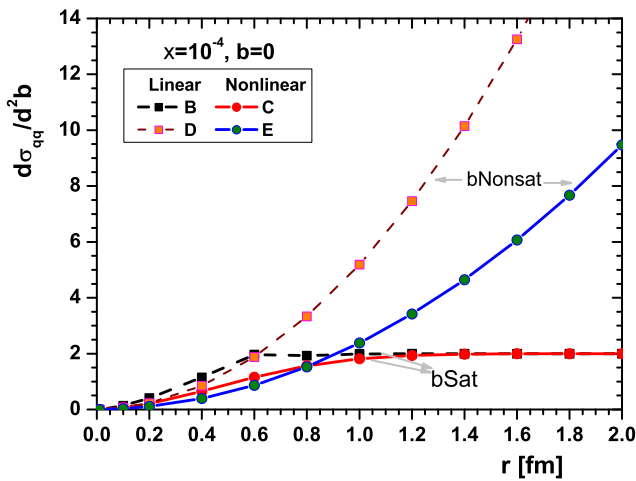}
\caption{Comparison of linear and nonlinear corrections to the
$d\sigma_{\mathrm{dip}}/d^{2}b$ for the IP-Sat (Eq.(6)) as well as
IP Non-Sat (Eq.(11)) versus $r$ for $x=10^{-4}$ and
$b=0$.}\label{Fig7}
\end{figure}
A comparison of the resulting $d\sigma_{\mathrm{dip}}/d^{2}b$
according to the linear as well as nonlinear behavior of the
dipole cross section for $b=0$ at $x=10^{-4}$ presented in Fig.7.
The resulting dipole cross-sections  in linear and nonlinear
corrections are shown in Fig. 7. We observe that, in this figure,
the nonlinear corrections suppress the behavior of the large
dipoles in the IP Non-Sat model. Indeed, this behavior tamed at
large $r$ for $b=0$ where with the increase $r$, $\mu$ decreases
to the value of $\mu_{0}$. We also note that adding nonlinear
corrections to the IP-Sat model decreases the dipole cross-section
for $0.2~\mathrm{fm}<r<1~\mathrm{fm}$ at $b=0$. The linear and
nonlinear corrections to the dipole cross section to the IP-Sat
model reach the saturation plateau at $r>1~\mathrm{fm}$.\\
\begin{figure}
\includegraphics[width=0.55\textwidth]{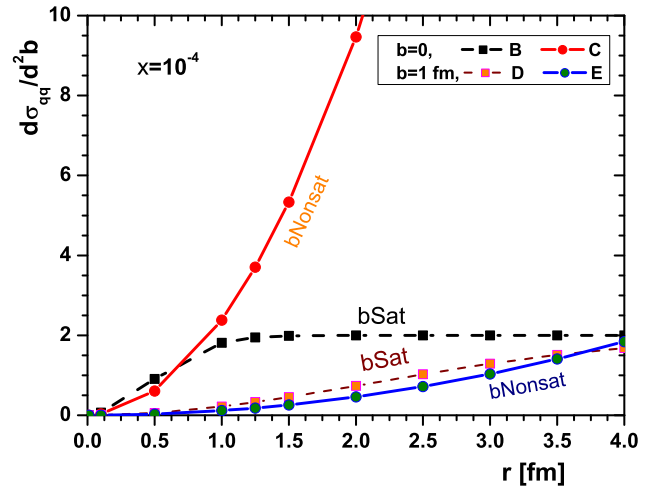}
\caption{Comparison of nonlinear corrections to the
$d\sigma_{\mathrm{dip}}/d^{2}b$ for the IP-Sat (Eq.(6)) as well as
IP Non-Sat (Eq.(11)) versus $r$ at $x=10^{-4}$ for $b=0$ and
$b=1~\mathrm{fm}$, respectively.}\label{Fig8}
\end{figure}
It is interesting to increase the impact parameter from $b=0$ to
$1~\mathrm{fm}$ for the nonlinear behavior of the dipole cross
sections to the IP Non-Sat in Fig.8. The proton dipole cross
section at different impact parameters with and without nonlinear
corrections are shown in Fig.8 for the IP-Sat as well as IP
Non-Sat versus $r$ at $x=10^{-4}$. The IP-Sat and IP Non-Sat
dipole cross sections are very similar in the range
$0{\leq}r{\leq}4~\mathrm{fm}$ for $b=1~\mathrm{fm}$. Consequently,
for large impact parameter sizes the distributions are almost
similar but they differ significantly as $b$ becomes small due to
the nonlinear corrections. Indeed, the nonlinear corrections
become stronger at larger impact parameters for the IP Non-Sat
model [3]. In Fig.9 we show the IP Non-Sat to IP-sat cross-section
ratios as a function of $r$  for $x=10^{-4}$. We depict the ratio
as a function of $r$ for $b=0$ and $1\mathrm{fm}$. Note that the
ratio increases much faster as a function of $r$  for $b=0$ than
for $b=1\mathrm{fm}$. We further note that at larger $r$, the
ratio remains near almost unity for $b=1\mathrm{fm}$. The large
difference between IP Non-Sat and IP-Sat comes from the decreases
in the impact parameter values.\\
\begin{figure}
\includegraphics[width=0.55\textwidth]{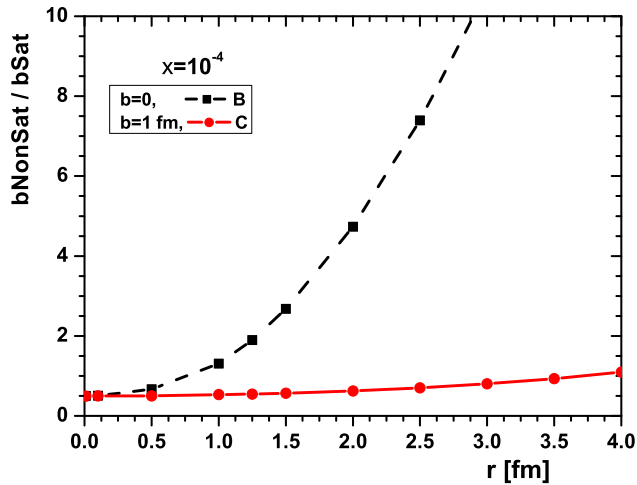}
\caption{The nonlinear corrections to the ratio IP Non-sat/IP-Sat
(Eq.(11)/Eq.(6)) of the $d\sigma_{\mathrm{dip}}/d^{2}b$ as a
function $r$ at $x=10^{-4}$ for $b=0$ and $1~\mathrm{fm}$,
respectively.}\label{Fig9}
\end{figure}

\subsection{V. Conclusions}

In this paper we proposed a modification of the saturation model
which takes into account the GLR-MQ evolution of the gluon
distribution. We have presented a certain theoretical model at LO
approximation to describe the color dipole cross sections based on
the Laplace transforms method at small values of $x$ (the Bjorken
variable $x$ is fixed to be $x=10^{-4}$). We have used a nonlinear
correction to the dipole cross sections from a parametrization of
the proton structure function with a rescaled variable $m_{c}$.
The nonlinear corrections to the dipole cross sections in the
description of inclusive and diffractive DIS at small $x$,
according to the saturation scale and geometric scaling, are
consistent with analytical saturation models in a wide range of
$r$ and $rQ_{s}$, respectively. We find that the ratio
$\sigma_{\mathrm{dip}}/\sigma_{0}$ due to the DGLAP improved
saturation model is consistent with the GBW saturation model,
although the nonlinear corrections to this ratio with respect to
the GLR-MQ improved saturation model is consistent with the CGC
saturation model especially in the range $0.05<r{\leq}1$. In the
simplest case of the $q\overline{q}$ system for the ratio
$\sigma^{2}_{\mathrm{dip}}/\sigma^{2}_{0}$ in the diffractive
processes, the linear and nonlinear corrections show good
agreement with the GBW and CGC models in a wide range of $r$ and
$rQ_{s}$. The linear and nonlinear corrections to the ratio
$\sigma^{2}_{\mathrm{dip}}/\sigma^{2}_{0}$ in the diffractive
processes due to the component $q\overline{q}g$
 deviates from the GBW and CGC models, because the
 $q\overline{q}g$ system goes beyond the saturation models.\\
We developed nonlinear corrections to the impact parameter
dependent dipole cross sections, $d\sigma_{\mathrm{dip}}/d^{2}b$.
The nonlinear corrections to the IP-Sat model are comparable with
the b-CGC model in a wide range of the impact parameter $b$ and
the dipole size $r$. The linear and nonlinear corrections
considered in the IP-Sat and IP Non-Sat models for the impact
parameters $b=0$ and $b=1~\mathrm{fm}$ in the range
$0{\leq}r{\leq}4$. The behavior of the nonlinear corrections to
the IP Non-Sat model tamed in a wide range of $r$. This behavior
for the IP-Sat model shows that the dipole cross section saturated
early for $b=0$ in comparison with  $b=1~\mathrm{fm}$ for
$r>1~\mathrm{fm}$. The nonlinear corrections to the IP-Sat and IP
Non-Sat models show that those behaviors are comparable in the
range $0{\leq}r{\leq}4$ for the impact parameter
$b=1~\mathrm{fm}$.\\
In conclusion, by considering the statistical errors due to the
effective parameters, the nonlinear corrections to the dipole
cross sections give a reasonable data description in comparison
with the other models. Indeed, the GLR-MQ improved saturation
model tames the DGLAP improved model behavior when the results
compared to models described based on the recombination of
gluons at low $x$.\\

\subsection{ACKNOWLEDGMENTS}
The author is grateful to Razi University for the financial
 support of this project.\\

%

\section{References}
1. K.Golec-Biernat and  M.W$\ddot{\mathrm{u}}$sthoff, Phys. Rev. D
{\bf59},  014017 (1999); K. Golec-Biernat and S.Sapeta, JHEP
{\bf03}, 102 (2018).\\
2. J.Bartels, K.Golec-Biernat and H.Kowalski, Phys. Rev. D{\bf66},
014001 (2002).\\
3. B.Sambasivam, T.Toll and T.Ullrich, Phys.Lett.B {\bf803}, 135277 (2020).\\
4. J.R.Forshaw and G.Shaw, JHEP {\bf12},
052 (2004).\\
5. E.Iancu, A.Leonidov and L.McLerran, Nucl.Phys.A {\bf692}, 583
(2001); Phys.Lett.B {\bf510}, 133 (2001); E.Iancu,K.Itakura and
S.Munier, Phys.Lett.B {\bf590}, 199
(2004).\\
6. Yu.L.Dokshitzer, Sov.Phys.JETP{\bf46}, 641 (1977); G.Altarelli
and G.Parisi, Nucl.Phys.B {\bf126}, 298 (1977); V.N.Gribov and
L.N.Lipatov, Sov.J.Nucl.Phys. {\bf15},
438 (1972).\\
7. A. H. Mueller and J. Qiu, Nucl. Phys. B{\bf268}, 427 (1986); L.
V. Gribov, E. M. Levin and M. G. Ryskin, Phys.
Rep.{\bf100}, 1 (1983).\\
8. N. N. Nikolaev and W. Sch$\ddot{a}$fer, Phys. Rev. D{\bf74}, 014023 (2006);
 N.N. Nikolaev, B.G. Zakharov, Z. Phys. C {\bf49}, 607 (1991); Z. Phys. C {\bf53}, 331 (1992)\\
9. Qinq-Dong Wu et al., Chin.Phys.Lett. {\bf33}, 012502 (2016).\\
10. H.Kowalski and D.Teaney, Phys. Rev. D{\bf68}, 114005 (2003);
J.L.Abelleira Fernandez et al. [LHeC Study Group], J. Phys. G
{\bf39}, 075001 (2012); P.Agostini et al. [LHeC
Collaboration and FCC-he Study Group], J. Phys. G {\bf48},  110501 (2021).\\
11. A.H.Rezaeian, M.Siddikov, M. Van de Klundert, and R.
Venugopalan, Phys.Rev.D {\bf87}, 034002 (2013); A.H.Rezaeian and
I.Schmidt, Phys. Rev. D{\bf88}, 074016 (2013).\\
12. H. Kowalski, T. Lappi, and R. Venugopalan, Phys.Rev.Lett.
{\bf100}, 022303 (2008); H. Kowalski, T. Lappi, C. Marquet, and R.
Venugopalan, Phys.Rev.C {\bf78}, 045201 (2008).\\
13. A.H. Mueller, Nucl. Phys. B {\bf335}, 115 (1990).\\
14. L. McLerran, R. Venugopalan, Phys. Rev. D {\bf49}, 2233
(1994).\\
15. J.Bartels, K.Golec-Biernat and H.Kowalski, Acta Phys.Polon.B
{\bf33}, 2853 (2002);  K.Golec-Biernat and S. Sapeta, Phys. Rev.
D{\bf74}, 054032 (2006).\\
16. G.Watt and H.Kowalski, Phys. Rev.
D{\bf78}, 014016 (2008).\\
17. V.S.Fadin, E.A.Kuraev and L.N.Lipatov, Phys.Lett.B
\textbf{60}, 50(1975); L.N.Lipatov, Sov.J.Nucl.Phys. \textbf{23},
338(1976); I.I.Balitsky and L.N.Lipatov, Sov.J.Nucl.Phys.
\textbf{28}, 822(1978).\\
18. K.Golec-Biernat, J.Phys.G {\bf28}, 1057 (2002); Acta Phys.Polon.B {\bf33}, 2771 (2002).\\
19. Y.S.Jeong, C.S.Kim, M.V.Luu and M.H.Reno, JHEP {\bf11}, 025
(2014).\\
20. G.R.Boroun, Eur.Phys.J.C {\bf82}, 740 (2022).\\
21. G.R.Boroun, Eur.Phys.J.Plus {\bf137}, 371 (2022);
 G.R.Boroun and B.Rezaei, Eur.Phys.
J.C {\bf81}, 851 (2021).\\
22. M.R.Pelicer et al., Eur.Phys.J.C {\bf79}, 9 (2019); M. Devee, J.K. Sarma, Eur. Phys. J. C {\bf74}, 2751(2014).\\
23. Edmond L. Berger, M. M. Block and Chung-I Tan, Phys. Rev.
Lett.
{\bf98}, 242001 (2007).\\
24. M.M.Block and L.Durand and D.W.Mckay, Phys.Rev.D {\bf77},
094003 (2008); M.M.Block and L.Durand, arXiv [hep-ph]:0902.0372.\\
25. J.Kwiecinski et al., Phys.Rev.D {\bf42}, 3645 (1990).\\
26. H. Abramowicz et al. (H1 and ZEUS Collaborations), Eur. Phys.
J. C {\bf78}, 473 (2018).\\


\end{document}